\documentclass[a4paper,11pt]{article}
\usepackage[a4paper, total={17cm, 25cm}]{geometry}
\usepackage{fancyhdr}
\usepackage{authblk}
\usepackage{graphicx}

\fancypagestyle{plain}{
\fancyhf{} 
\fancyhead[RH]{\thepage}
\fancyfoot[CF]{28th Texas Symposium on Relativistic Astrophysics \\ Geneva, Switzerland -- December 13-18, 2015}

}
\begin{document}

\title{\bf High-resolution tSZ cartography of clusters of galaxies with NIKA at the IRAM 30-m telescope}

\author[1]{F.~Mayet}
\author[1,13]{R.~Adam}
\author[2]{A.~Adane}
\author[3]{P.~Ade}
\author[4]{P.~Andr\'e}
\author[4]{M. Arnaud}
\author[4]{I. Bartalucci}
\author[5]{A.~Beelen}
\author[6]{A.~Beno\^it}
\author[3]{A.~Bideaud}
\author[7]{N.~Billot}
\author[1]{G.~Blanquer}
\author[6]{N.~Boudou}
\author[1]{O.~Bourrion}
\author[6]{M.~Calvo}
\author[1]{A.~Catalano}
\author[2]{G.~Coiffard}
\author[1]{B.~Comis}
\author[8]{A.~Cruciani}
\author[9]{F.-X.~D\'esert}
\author[3]{S.~Doyle}
\author[6]{J.~Goupy}
\author[5]{B. Hasnoun}
\author[7]{I.~Hermelo}
\author[7]{C.~Kramer}
\author[10]{G. Lagache}
\author[2]{S.~Leclercq}
\author[1]{J.~F.~Mac\'ias-P\'erez}
\author[3,14]{P.~Mauskopf}
\author[6]{A.~Monfardini}
\author[5]{F.~Pajot}
\author[1]{L.~Perotto}
\author[11,12]{E.~Pointecouteau}
\author[9]{N.~Ponthieu}
\author[4]{G.~W.~Pratt}
\author[4]{V.~Rev\'eret}
\author[1]{A.~Ritacco}
\author[4]{L.~Rodriguez}
\author[1]{F.~Ruppin}
\author[2]{K.~Schuster}
\author[7]{A.~Sievers}
\author[6]{S.~Triqueneaux}
\author[3]{C.~Tucker}
\author[2]{R.~Zylka}

\affil[1]{Laboratoire de Physique Subatomique et de Cosmologie,
Universit\'e Grenoble Alpes, CNRS/IN2P3, 53, avenue des Martyrs, Grenoble, France}
\affil[2]{Institut de RadioAstronomie Millim\'etrique (IRAM), Grenoble, France}
\affil[3]{Astronomy Instrumentation Group, University of Cardiff, UK}
\affil[4]{Laboratoire AIM, CEA/IRFU, CNRS/INSU, Université Paris Diderot, CEA-Saclay, 91191 Gif-Sur-Yvette, France} 
\affil[5]{Institut d'Astrophysique Spatiale (5), CNRS and Universit\'e Paris Sud, Orsay, France}
\affil[6]{Institut N\'eel, CNRS and Universit\'e de Grenoble, France}
\affil[7]{Institut de RadioAstronomie Millimetrique (IRAM), Granada, Spain}
\affil[8]{Dipartimento di Fisica, Sapienza Universit\`a di Roma, Piazzale Aldo Moro 5, I-00185 Roma, Italy}
\affil[9]{Institut de Plan\'etologie et d'Astrophysique de Grenoble (8), CNRS and Universit\'e Grenoble Alpes, France}
\affil[10]{Aix Marseille Universit\'e, CNRS, LAM (Laboratoire d'Astrophysique de Marseille) UMR 7326, 13388, Marseille, France}
\affil[11]{Universit\'e de Toulouse, UPS-OMP, Institut de Recherche en Astrophysique et Plan\'etologie (IRAP), Toulouse, France}
\affil[12]{CNRS, IRAP, 9 Av. colonel Roche, BP 44346, F-31028 Toulouse cedex 4, Franc}
\affil[13]{Laboratoire Lagrange, Universit\'e C\^ote d'Azur, Observatoire de la C\^ote d'Azur, CNRS, 
Blvd de l'Observatoire, CS 34229, 06304 Nice Cedex 4, France} 
\affil[14]{School of Earth and Space Exploration and Department of Physics, Arizona State University, Tempe, AZ 85287, USA}


\maketitle

\begin{abstract}
The thermal Sunyaev-Zeldovich effect (tSZ) is a powerful probe to study clusters of galaxies and  
is complementary with respect to X-ray, lensing or optical observations.
Previous arcmin resolution tSZ observations ({\it e.g.} SPT, ACT and Planck) only enabled 
detailed studies of the intra-cluster medium morphology for low redshift clusters ($z < 0.2$). 
Thus, the development of precision cosmology with clusters 
requires high angular resolution observations to extend the understanding 
of galaxy cluster towards high redshift. 
NIKA2 is a wide-field (6.5 arcmin field of view) dual-band camera, operated at 
$100 \ {\rm mK}$ and containing $\sim 3300$ KID (Kinetic Inductance Detectors), 
designed to observe the millimeter sky at 150 and 260 GHz, 
with an angular resolution of 18 and 12 arcsec respectively. 
The NIKA2 camera   has been installed on the IRAM 30-m telescope 
(Pico Veleta, Spain) in September 2015. The NIKA2 tSZ observation program 
will allow us to observe a large sample of clusters (50) at redshift ranging  between 
0.5 and 1. As a pathfinder for NIKA2, several clusters of galaxies have been observed at 
the IRAM 30-m telescope with the NIKA prototype to cover the various configurations and observation 
conditions expected for NIKA2.
\end{abstract}


\section{High-resolution tSZ cartography of clusters}

Clusters of galaxies constitute powerful tools to study cosmology as they are the largest gravitationally-bound 
objects in the Universe. For instance, cosmological parameters have been constrained 
by the Planck collaboration using number counts as a function of 
redshift for a sample of galaxy clusters identified by their thermal Sunyaev Zel'dovich (tSZ) effect \cite{Ade:2013lmv,Ade:2015fva}. 
However, this estimation of the  amplitude of density fluctuations $\sigma_8$ and of the matter density $\Omega_M$, 
obtained with clusters of galaxies, are in tension 
with the values  obtained with  primary CMB anisotropies \cite{Ade:2013zuv}. 
The use of clusters of galaxies for cosmological studies requires to  translate cluster observables into mass estimates. However, 
non-gravitational processes can induce dispersion and biases. Also, at high redshift a departure from the 
hydrostatic equilibrium hypothesis is expected. 
Moreover, an X-ray based calibration of the tSZ-mass scaling law is needed in order 
to draw cosmological conclusions. As outlined in \cite{Ade:2013lmv}, possible biases 
in the scaling law relation and the halo mass function dominate the 
statistical uncertainties from the Planck cluster sample. Hence, improving the precision of cluster mass
calibration is required to strengthen the use of clusters of galaxies for cosmological studies.
A high-resolution cartography of clusters of galaxies is also needed to fully understand  
the intra-cluster medium properties and to evaluate the  hydrostatic bias. Thus, the development 
of precision cosmology with clusters requires   high-resolution multi-probe studies 
(X-ray, tSZ, lensing, optical) up to high redshift ($z \simeq 1$).\\

The tSZ effect \cite{Sunyaev:1972,Sunyaev:1980vz,Birkinshaw:1998qp} produces the distortion of the electromagnetic spectrum of the cosmic microwave background 
via an inverse Compton scattering with hot electrons in the intra-cluster medium. 
As the tSZ surface brightness does not suffer from redshift dimming, 
it enables the study of clusters up to high redshift. In the last few years, 
the tSZ effect has been used to detect clusters of galaxies. In particular, the South Pole Telescope 
(SPT)~\cite{Bleem:2014iim}, the Atacama Cosmology Telescope 
(ACT)~\cite{Hasselfield:2013wf} and the Planck Satellite \cite{Ade:2015gva,Ade:2013skr}  have produced 
tSZ-selected cluster catalogues, containing thousands of objects, at arcmin resolution. 
tSZ surveys are considered as major tools of multi-probe studies of clusters of galaxies. In particular, the complementarity with X-ray
surveys is highlighted by the fact that the X-ray surface 
brightness is related to the electronic density,  
$S_X   \propto \int n_e^2 \Lambda(T_e, Z) dl$, while the Compton parameter $y$, the tSZ signal, 
is proportional to the electronic pressure $P_{\mathrm{e}}$ integrated along the line 
of sight $y \propto \int P_{\mathrm{e}} dl$, thus giving information on shocks and merging events.

\section{The NIKA2 camera and the NIKA prototype}
The NIKA2 camera is a next-generation instrument for millimetre astronomy \cite{Monfardini11,Bourrion12,Calvo13,Monfardini13,Calvo16}. 
It is a KID-based camera operated at 100 mK that has been installed in September 2015 at the IRAM 30-m telescope (Pico Veleta, Spain).
NIKA2 observes the sky at 150 and 260 GHz with a wide field of view 
(6.5 arcmin) at high-angular resolution (nominally 18 and 12 arcsec, respectively), 
and state-of-art sensitivity (requirement 20 and 30 $\rm mJy.s^{1/2}$, respectively). NIKA2 has also polarisation capabilities at 260 
GHz. 
The NIKA camera is a prototype of NIKA2 that has been operated at the IRAM 30-m telescope from 2012 to 2015. The field of view is smaller
(1.8 arcmin) due to the reduced number of KIDs (356). The performance of the 
NIKA camera at the IRAM 30 m telescope is described in \cite{Catalano:2014nml}.\\

The NIKA2 camera is  well suited for high-resolution tSZ observations of  cluster of galaxies because:
\begin{enumerate}
\item It is a dual-band camera operating at frequencies (150 and 260 GHz) for which the tSZ signal is expected to be negative and slightly
positive respectively. 

\item NIKA2 is made of arrays of thousands of highly sensitive KIDs. In particular,  the  
sensitivity in Compton parameter units is expected to be of $1.13 \times 10^{-4}$ per hour and per beam. 
This will allow us to obtain reliable tSZ mapping at high signal-to-noise ratio in a few hours per cluster.

\item NIKA2 coupled to the IRAM 30-m telescope should allow us to map clusters of galaxies to a resolution of 
typically 12 to 18 arcsec for a 6.5 arcmin diameter FOV, which is well adapted for medium and high redshift clusters.
\end{enumerate}

%
\begin{figure*}[t]
\begin{center}
\includegraphics[scale=0.48,angle=0]{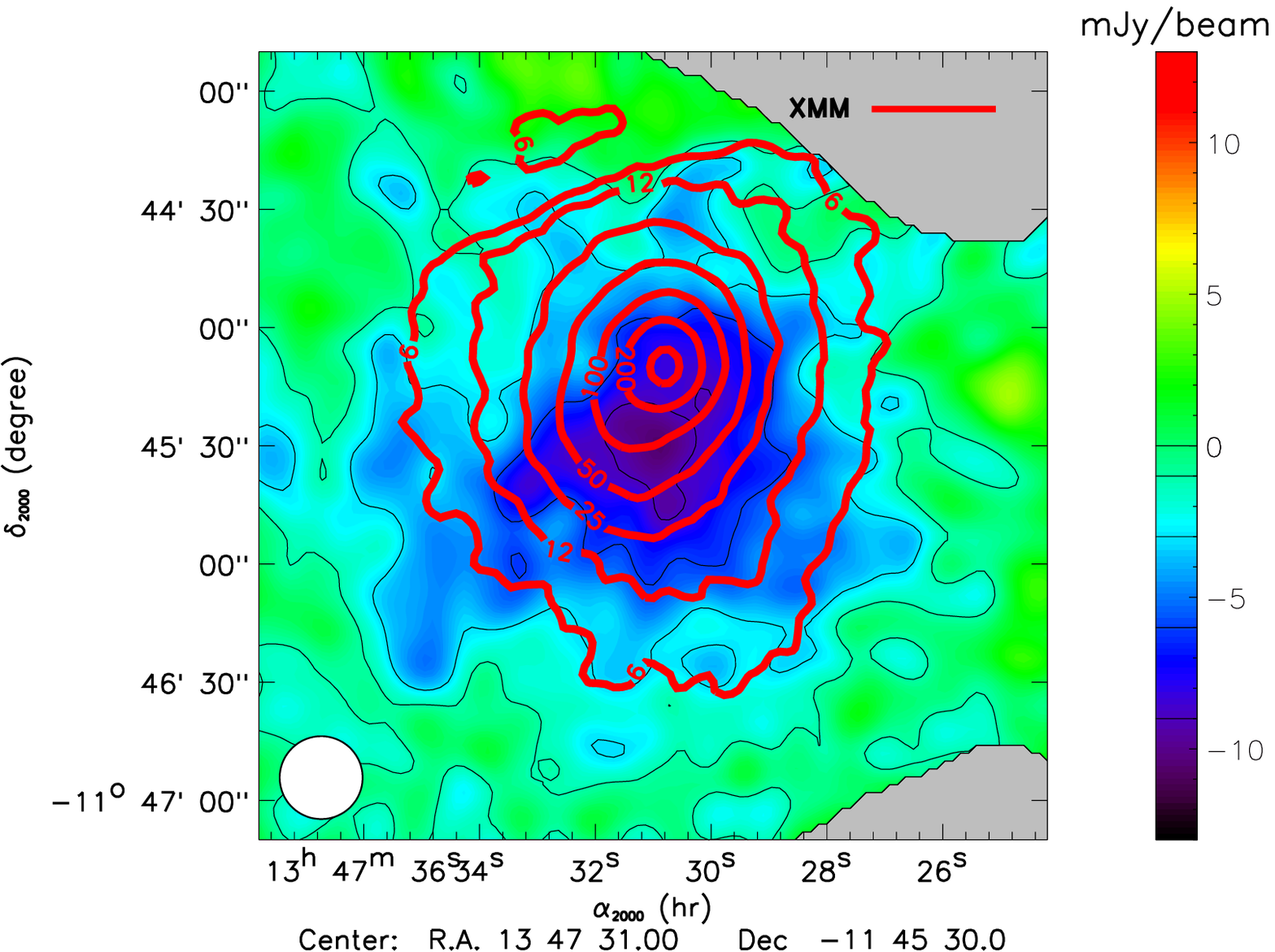}
\includegraphics[scale=0.45,angle=0]{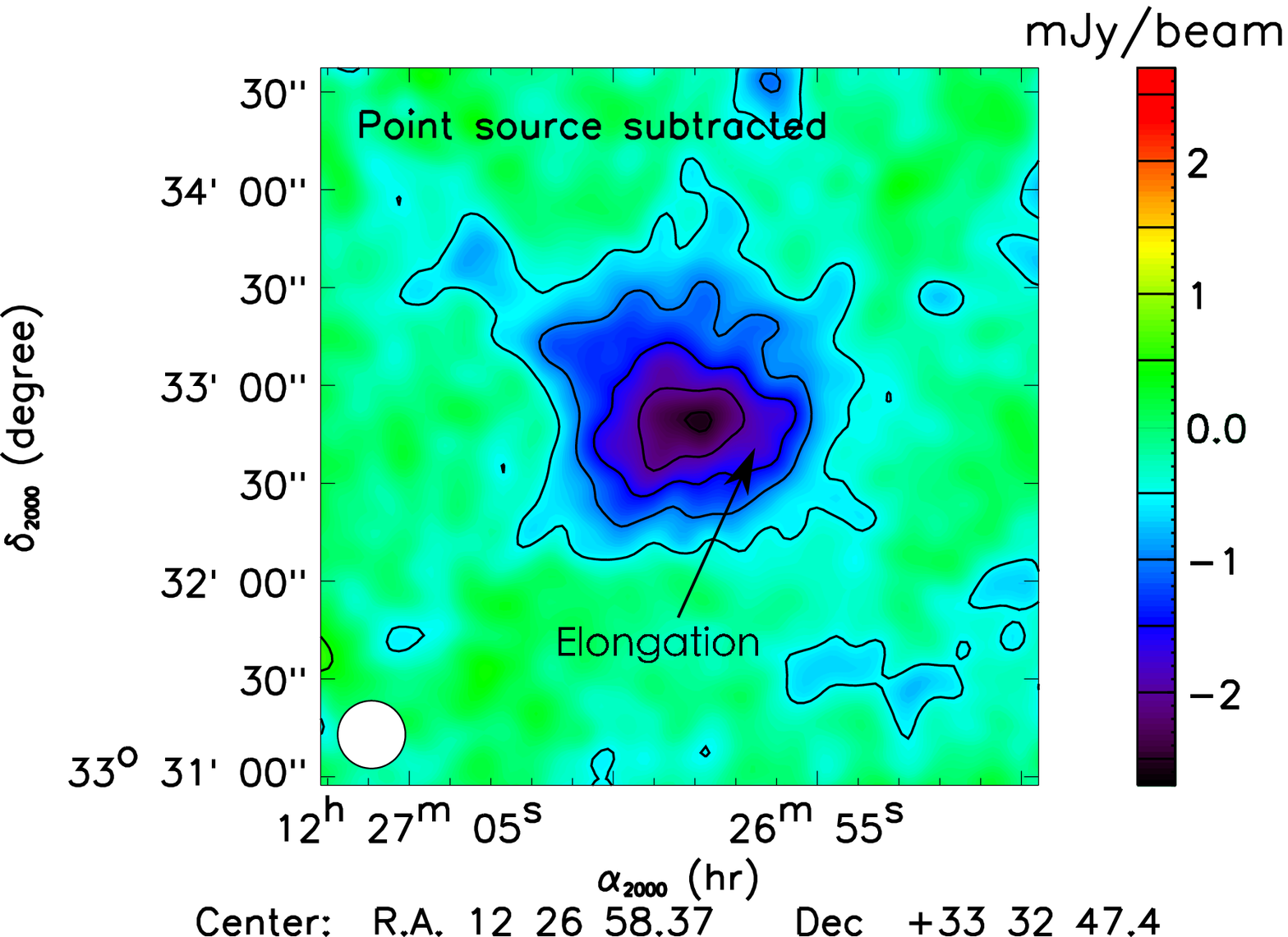}
\caption{Left: RX J1347.5-1145 as observed by NIKA in 2012 at 140 GHz \cite{Adam:2013ufa}. Right: CLJ1226.9+3332 
as observed by NIKA in 2014 at 150 GHz \cite{Adam:2014wxa}.}   
\label{fig:maps}
\end{center}
\end{figure*}

\section{tSZ cartography of clusters of galaxies with NIKA}
The NIKA prototype has been used as a pathfinder for NIKA2, to demonstrate the possibility to use large arrays of KIDs in millimeter astronomy, 
to validate the observation strategy as well as the data analysis. Concerning tSZ science, several clusters of galaxies have been observed during the three NIKA campaigns at the IRAM 30-m telescope.
They have been chosen in order to cover the various configurations and observation 
conditions expected for NIKA2.\\

RX J1347.5-1145 is the most luminous X-ray cluster known to date. It is also known as a particularly bright tSZ source 
that has been observed 
by {\it e.g.} Diabolo \cite{pointecouteau},  Mustang \cite{Mason:2009fw} or Carma \cite{Plagge:2012nr}. It has been chosen as a first target with NIKA and has been observed at the IRAM 30-m
telescope in 2012 during 5.5 hours. The data analysis of this cluster includes a dual-band atmospheric 
noise removal which consists in using the 260 GHz data-set as an atmospheric template since 
the expected tSZ signal is small at this frequency. It leads to a map of this cluster at 140 GHz presented on 
Fig.~\ref{fig:maps} (left) and compared with the XMM contours \cite{Gitti:2004ih}. It can be noticed that the tSZ peak is shifted 
toward the south-east with respect to the X-ray peak, which is well aligned on the central AGN.  As the tSZ flux is proportional to the
electronic pressure integrated along the line of sight, this  south-east extension indicates an overpressure corresponding to the expected shock caused by the ongoing merger. It   
is also observed in the radial flux profile  and in the residual of the map with respect to the modeling of the relaxed part of the
cluster. These observations constitute the first tSZ observation ever performed with a KID-based camera and are a confirmation that RX J1347.5-1145 is an ongoing merger \cite{Adam:2013ufa}.\\

\begin{figure*}[t]
	\centering
	\includegraphics[scale=0.48]{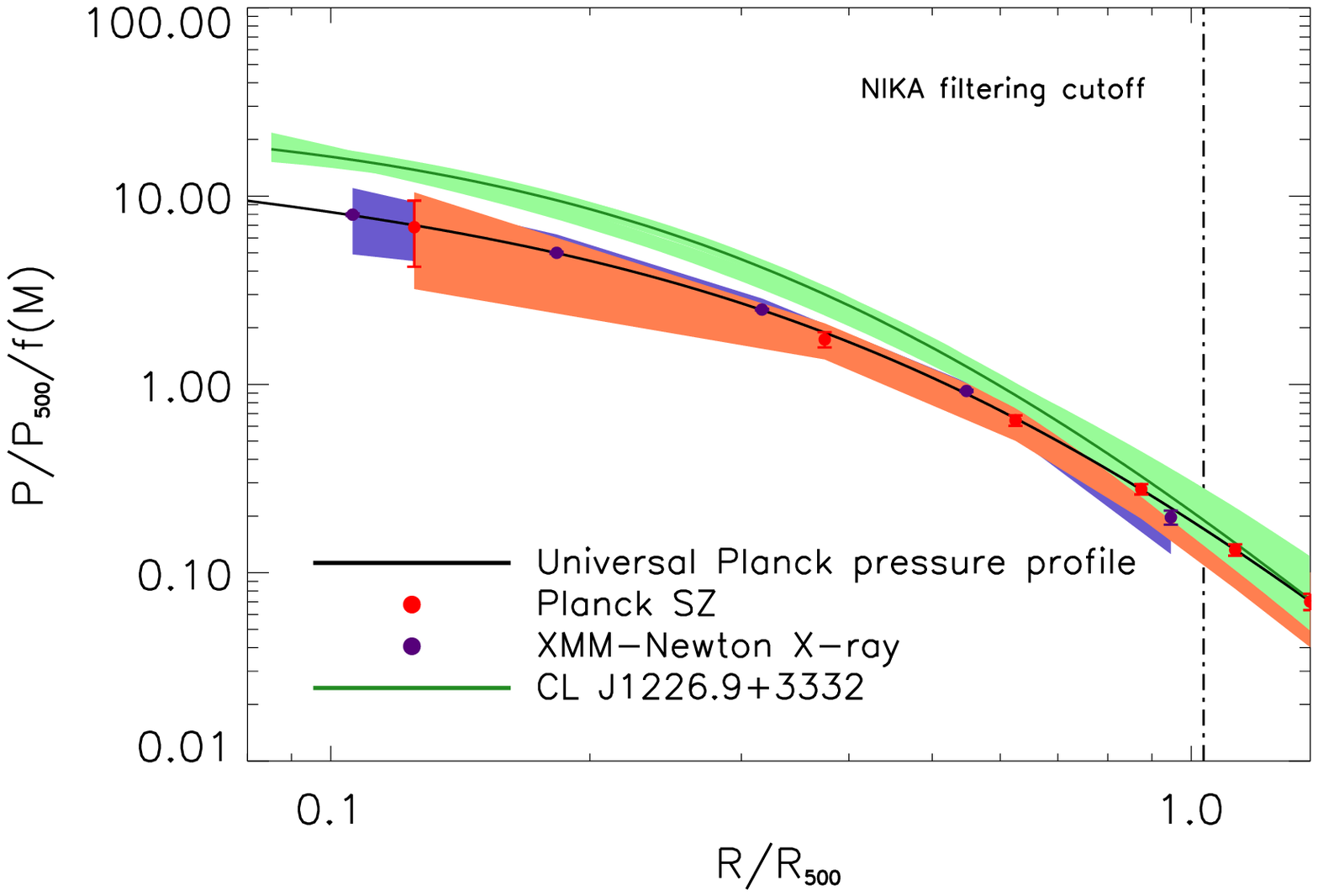}
	\includegraphics[scale=0.48]{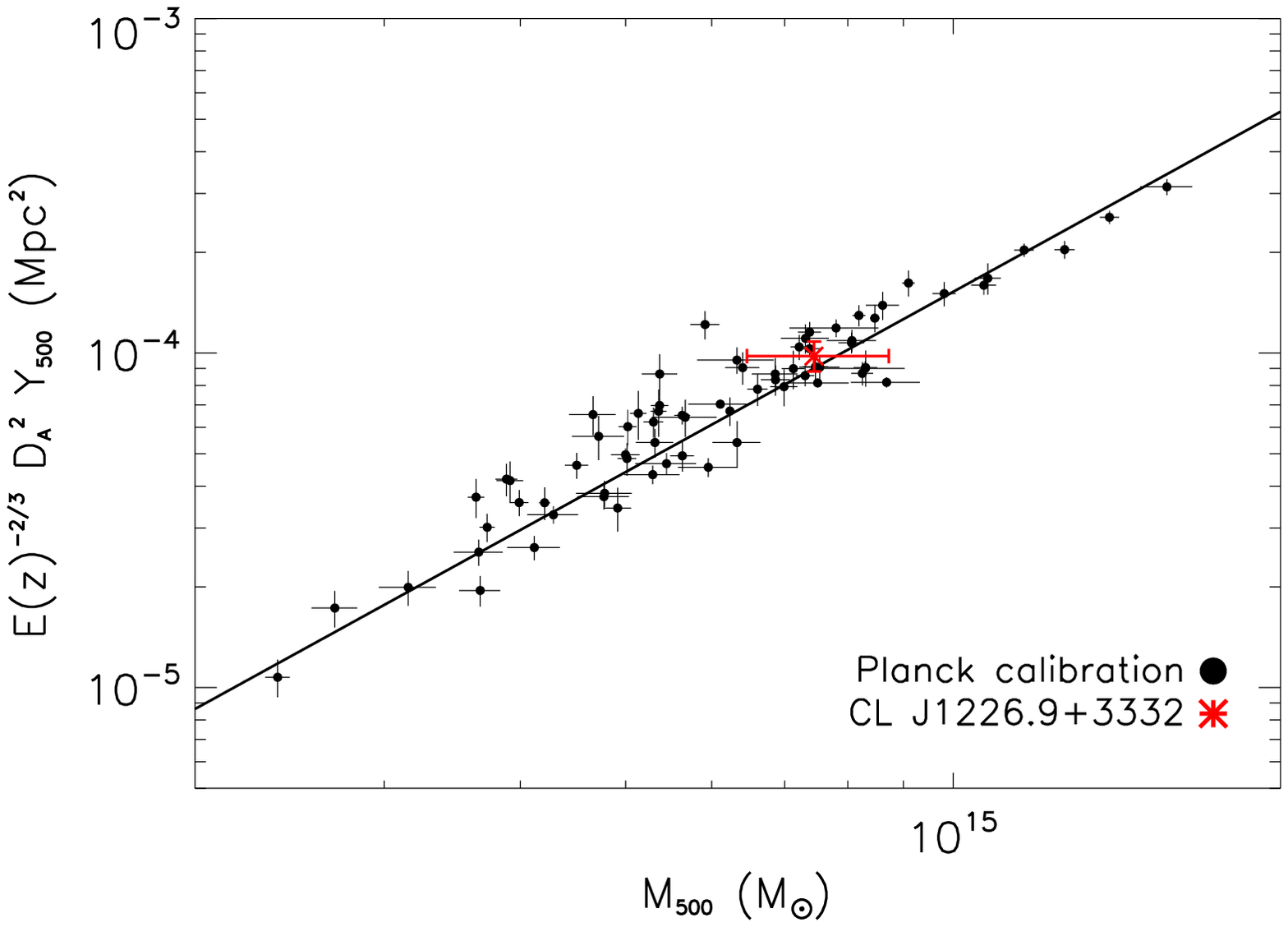}
	\caption{Left: Normalized pressure profile with the Planck Universal pressure profile (black line), the CL J1226.9+3332 profile from NIKA (green), the 
	 Planck average of  62 
	nearby clusters (red) and the stacked pressure profile derived from the 
	XMM data for the same sample (purple). Right: tSZ scaling relation, relating  the Compton integrated parameter  $Y_{500}$ and  the cluster mass $M_{500}$.  CL J1226.9+3332, at $z = 0.89$, 
	is given by the red star. The Planck best-fit is presented as a black line, with the data points of the 71 clusters. 
	Figures from \cite{Adam:2014wxa}}
        \label{fig:scaling_law}
	\end{figure*}

CL J1226.9+3332 is a relaxed  high-redshift ($z=0.89$) cluster that has been 
observed by NIKA in 2014, during 7.8 hours \cite{Adam:2014wxa}. The 150 GHz map, presented on 
Fig.~\ref{fig:maps} (right), shows that CL~J1226.9+3332 is relaxed on large scales with  
a disturbed core. The 260 GHz channel has been used  to identify point-source contamination. A point-source subtraction method has been
used to correct for the induced deformation at 150 GHz. 
NIKA data have been combined with Planck tSZ data and X-ray from Chandra \cite{Cavagnolo:2009kn} within the framework of a multi-probe
analysis to study the thermodynamic properties of the intra-cluster medium, {\it i.e.} the pressure, density, temperature, and entropy radial 
distributions. It illustrates the possibility of measuring these quantities with a small integration time, even at high redshift, and without X-ray spectroscopy. The pressure profile of this high-redshift cluster, obtained with NIKA, has been compared 
with the profile averaged  
over 62 nearby clusters obtained by the Planck collaboration \cite{planck.pressure}, see Fig. \ref{fig:scaling_law} (left). 
Both have been normalized to account for the mass and redshift dependance. While the normalized pressure profile of 
CL J1226.9+3332, derived from NIKA data, is higher than the average one, it remains compatible within error bars. 
The tSZ scaling relation, relating  the Compton integrated parameter  $Y_{500}$ and  the cluster mass $M_{500}$, is presented on Fig. \ref{fig:scaling_law} (right). 
The cluster CL J1226.9+3332, at $z = 0.89$, is consistent with the Planck best-fit scaling relation \cite{Ade:2013lmv} obtained with  71 clusters, with a  mean redshift of 0.195 and a maximum of 0.447. 
While no conclusion can be drawn from the comparison with a single high-redshift cluster, it illustrates  the interest of the forthcoming 
NIKA2 tSZ large program, which is dedicated to a follow-up of Planck-discovered clusters,  to study non-standard redshift evolution of cluster pressure profile.\\

\begin{figure*}[t]
	\centering
	\includegraphics[scale=0.4]{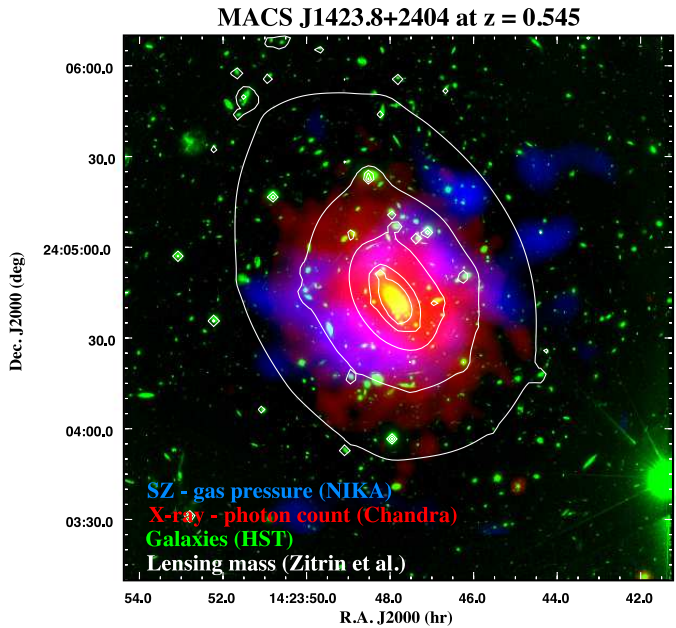}
	\caption{Multi-probe  image of
MACS J1423.8+2404, with tSZ at 150 GHz as seen by NIKA, X-ray as 
seen by Chandra, lensing \cite{Zitrin:2010rn} and galaxies as seen by the Hubble Space Telescope \cite{Postman:2011hg}. 
From \cite{Adam:2015bba}.}
        \label{fig:multiprobe}
	\end{figure*}

The observation of MACS J1423.8+2404 by NIKA has been used to show the impact of contamination 
from infrared and radio point sources. Indeed, most clusters host submillimeter and/or radio point sources that can 
significantly affect the reconstructed tSZ signal, in particular at frequency below 217 GHz where the negative 
tSZ signal can be compensated. This cluster has been observed in 2014 and the results have been obtained with only 
1.47 hours of on-target data~\cite{Adam:2015bba}. The map obtained at 150 GHz presents a $4.5\sigma$ detection at the tSZ peak. 
19 point sources have been identified in the $4 \times 4 \ {\rm arcmin}^2$ field around MACS J1423.8+2404. 
Ancillary data have been combined with NIKA data 
to study the SED of the submillimeter (Herschel) and radio (SZA, OVRO/BIMA, VLA and NVSS) point source contaminants, 
see Refs. in \cite{Adam:2015bba}. A multi-probe study of the intracluster medium has been performed by combining data from NIKA, Planck, XMM-Newton and Chandra. 
As an illustration, Fig.~\ref{fig:multiprobe} presents a composite multi-probe overview image of
MACS J1423.8+2404, with tSZ at 150~GHz as seen by NIKA, X-ray as seen by Chandra, see Ref.~\cite{Adam:2015bba}, lensing 
\cite{Zitrin:2010rn} and galaxies as seen by the Hubble Space Telescope \cite{Postman:2011hg}. 
This paper has shown that an accurate removal of the point-source contamination is required in order 
to set strong constraints on the central pressure distribution.\\

Furthermore, we have observed two clusters initially discovered by Planck via the tSZ effect, 
PSZ1G045.85+57.71 and PSZ1G046. A detailed analysis of these clusters 
is ongoing and will complete the NIKA pilote study in preparation of the NIKA2 tSZ program.

\section{The NIKA2 tSZ large program}
The NIKA2 tSZ large program is a follow-up of Planck-discovered clusters that has been proposed as one of the 
Large Programs of the NIKA2 Guaranteed time. 50 clusters will be observed, with redshift  up to $z=1$, selected from the 
Planck \cite{Ade:2015gva,Ade:2013skr} and ACT catalogs \cite{Hasselfield:2013wf}. 
The expected sensitivity of NIKA2, based on what was achieved by NIKA, 
will allow us to obtain reliable tSZ detection and mapping of clusters of galaxies in 
only few hours (1 to 5 hours). We have formed a representative cluster sample for 
redshift evolution and cosmological studies, with redshift bins presenting an 
homogeneous coverage in cluster mass as reconstructed from the integrated Compton parameter.  
The NIKA2 data will be complemented with ancillary data including X-ray, optical and radio observations. 
The full dataset, NIKA2 plus ancillary, will lead to 
significant improvements on the use of clusters of galaxies 
to draw cosmological constraints and in particular on the matter 
distribution and content of the Universe. 
The main objectives of the project are:
\begin{itemize}
\item the study of the redshift evolution of cluster pressure profiles up to high redshift ($z=1$), 
\item the understanding of cluster morphology at high redshift 
(merging events, departure from spherical symmetry, cooling processes), 
\item the detailed characterization of the physical properties of the cluster (temperature, entropy and mass radial profiles), within the
framework of a multi-probe analysis,
\item the reconstruction of scaling laws relating 
cluster global properties, the integrated Compton parameter and temperature for example, to their mass.
\end{itemize}


\section*{Acknowledgments}
We would like to thank the IRAM staff for their support during the campaigns. 
The NIKA dilution cryostat has been designed and built at the Institut N\'eel. 
In particular, we acknowledge the crucial contribution of the Cryogenics Group, and 
in particular Gregory Garde, Henri Rodenas, Jean Paul Leggeri, Philippe Camus. 
This work has been partially funded by the Foundation Nanoscience Grenoble, the LabEx FOCUS ANR-11-LABX-0013 and 
the ANR under the contracts "MKIDS", "NIKA" and ANR-15-CE31-0017. 
This work has benefited from the support of the European Research Council Advanced Grant ORISTARS 
under the European Union's Seventh Framework Programme (Grant Agreement no. 291294).
We acknowledge fundings from the ENIGMASS French LabEx (R. A. and F. R.), 
the CNES post-doctoral fellowship program (R. A.),  the CNES doctoral fellowship program (A. R.) and 
the FOCUS French LabEx doctoral fellowship program (A. R.).


\end{document}